\documentstyle[prb,aps,epsf,floats]{revtex}

\begin{document}

\renewcommand{\textfraction}{0.0}
\renewcommand{\floatpagefraction}{.7}
\setcounter{topnumber}{5}
\renewcommand{\topfraction}{1.0}
\setcounter{bottomnumber}{5}
\renewcommand{\bottomfraction}{1.0}
\setcounter{totalnumber}{5}
\setcounter{dbltopnumber}{2}
\renewcommand{\dbltopfraction}{0.9}
\renewcommand{\dblfloatpagefraction}{.7}

\draft

\twocolumn[\hsize\textwidth\columnwidth\hsize\csname@twocolumnfalse%
\endcsname

\title{Density Profiles in Random Quantum Spin Chains}

\author{Ferenc Igl\'oi}
\address{Research Institute for Solid State Physics, 
H-1525 Budapest, P.O.Box 49, Hungary}
\address{Institute for Theoretical Physics, Szeged University,
H-6720 Szeged, Hungary}

\author{Heiko Rieger}
\address{HLRZ c/o Forschungszentrum J\"ulich, 52425 J\"ulich, Germany}

\date{\today}

\maketitle

\begin{abstract}
We consider random transverse-field Ising spin chains and study the
magnetization and the energy-density profiles by numerically exact
calculations in rather large
finite systems ($L\le 128$). Using different boundary conditions
(free, fixed and mixed) the numerical data collapse to scaling functions,
which are very accurately described by simple analytic expressions.
The average magnetization profiles satisfy the Fisher-de
Gennes scaling conjecture and the corresponding scaling functions
are indistinguishable from those predicted by conformal invariance.
\end{abstract}


]

\newcommand{\bc}{\begin{center}}
\newcommand{\ec}{\end{center}}
\newcommand{\be}{\begin{equation}}
\newcommand{\ee}{\end{equation}}
\newcommand{\beqn}{\begin{eqnarray}}
\newcommand{\eeqn}{\end{eqnarray}}
\newcommand{\ba}{\begin{array}}
\newcommand{\ea}{\end{array}}

Every experimental system is geometrically constrained and therefore
has a surface, for which reason we have to discriminate between so called
{\it bulk} and {\it surface} properties. This is justified as long as
the correlation length is much smaller than the system size. However,
{\it at} a critical point it is more appropriate to describe the position
dependent physical properties of the system by density profiles rather
than bulk and/or surface observables.  For a number of universality
classes much is known about this spatially inhomogeneous behavior
\cite{review}, in particular in two dimensions, where conformal
invariance provides a powerful tool to study various geometries
\cite{2dreview}.

Not much is known about this issue for quantum systems with quenched
(i.e.\ time independent) {\it disorder}. Here one is confronted with a
possible quantum phase transition, i.e.\ a zero temperature transition
that is triggered by quantum rather than thermal fluctuations, as for
instance in random transverse Ising models
\cite{fisher,qsg,qsg2}. Their bulk properties have been studied quite
extensively by now. The aim of the present letter is to investigate
for the first time the above mentioned density profiles in a
geometrically constrained disordered system at a quantum phase
transition. In particular we study numerically the random transverse
field Ising chain and propose analytic expressions of the
magnetization and energy density profile for various boundary
conditions (b.c.).

In a critical system confined between two parallel plates, which are
a large but finite distance $L$ apart, the local densities $\langle
\Phi(r)\rangle$ such as the order parameter (magnetization) or the
energy density vary with the distance $l$ from one of the plates as a
smooth function of $l/L$. According to the scaling theory by Fisher
and de Gennes \cite{mefisher}:
\be
\langle \Phi(l)\rangle_{ab}=L^{-x_{\Phi}} F_{ab}(l/L)\;,
\label{one}
\ee
where $x_{\Phi}$ is the scaling dimension of the operator $\Phi$, while
$ab$ denotes the boundary conditions at the two plates. The scaling
function in (\ref{one}) has the asymptotic behavior:
\be
F_{ab}(l/L)=A\left[ 1+B_{ab}\left(l \over L \right)^d + \dots
\right] \quad {l \over L} \ll 1\;.
\label{two}
\ee
where the exponent in the
first correction term was confirmed by different methods
\cite{auyang,burkhardt1,cardy}.  It has been shown by Burkhardt and
Xue \cite{burkhardt2} and by Cardy \cite{cardy} that the $B_{ab}$
coefficients in (\ref{two}) and the $A_{ab}$ finite size correction
coefficients of the free energy as $A_{ab} L^{-d+1}$ are related to
each other: their ratio is universal and independent of the form of
the b.c.

Having the same type of b.c.\ at both plates the profile
$\langle \Phi(l)\rangle_{aa}=L^{-x_{\Phi}} f_{aa}(l/L)$
is reflection symmetric $f_{aa}(v)=f_{aa}(1-v)$ and according to
eqs.(\ref{one}) and (\ref{two}) $\lim_{v \to 0} f_{aa}(v) \sim
v^{-x_{\Phi}}$ . Consequently $\left[ f_{aa}(v) \right]^{-1/x_{\Phi}}$
can be expanded in a Fourier series
\cite{remark} and the profile is given by:
\be
\langle \Phi(l)\rangle_{aa}=L^{-x_{\Phi}} \left[\sum_{k=1}^{\infty} A_k
\sin {k \pi l \over L} \right]
^{-x_{\Phi}}
\label{four}
\ee
The Fourier expansion in (\ref{four}) has different convergence
properties in two- and three-dimensions due to the different parity of
the correction term in (\ref{two}).  While in three-dimensions
infinite terms are needed to reproduce the Fisher- de Gennes scaling
result in (\ref{two}), in two-dimensions one expect to obtain
satisfactory accuracy by the first few terms of the expansion. Indeed
for conformally invariant two-dimensional models only the first term
in the Fourier series in (\ref{four}) gives non-vanishing contribution
\cite{burkhardt1}:
\be
\langle \Phi(l)\rangle_{aa}=A \left[{L \over \pi} \sin \pi {l \over L} \right]
^{-x_{\Phi}}
\label{five}
\ee
Conformal invariance can be used further to predict the density profiles
with general b.c. In two-dimensions the profiles are
in the form \cite{burkhardt2}
\be
\langle \Phi(l)\rangle_{ab}= \left[{L \over \pi} \sin \pi {l \over L}
\right]^{-x_{\Phi}} G_{ab}\left( {l \over L} \right)
\label{six}
\ee
where the scaling function $G_{ab}(l/L)$ depends on the universality
class of the model and on the type of the b.c.  For
the Ising model the magnetization profile with free-fixed boundary
condition the scaling function is predicted as \cite{burkhardt2}:
\be
G_{f+}=B \left[ \sin { \pi l \over 2 L} \right]^{x_m^s}
\label{seven}
\ee
where $x_m^s=1/2$ is the scaling dimension of the surface
magnetization operator. Similar result is obtained for the $Q \le 4$
state Potts model \cite{burkhardt2} with the appropriate surface
scaling dimension in (\ref{seven}).

In the present Letter we consider the random transverse filed Ising chain
\be
\hat H=-\sum_l J_l \sigma_l^x \sigma_{l+1}^x -\sum_l h_l \sigma_l^z~~~,
\label{eight}
\ee
Here the $J_l$ exchange couplings and the $h_l$ transverse-fields are
independent random variables with distributions $\pi(J)$ and
$\rho(h)$, respectively and the $\sigma_l^x,\sigma_l^z$ are Pauli
matrices at site $l$. This Hamiltonian is the extreme anisotropic
limit \cite{kogut} of the layered two-dimensional Ising model as
introduced by McCoy and Wu \cite{mccoy1,mccoy2}.

The critical behavior of the random transverse-field Ising spin chain
in (\ref{eight}) has been investigated analytically
\cite{mccoy2,shankar,fisher} and numerically \cite{crisanti,young} in
several papers. Depending on the strength of the average value of the
transverse-field the system has two phases, which are separated by a
second order phase transition point located at \cite{shankar}
$\delta=\overline{\ln J}-\overline{\ln h}=0$.  Due to a broad
distribution of various physical quantities the typical and average
quantities of the system are generally different.  The scaling
dimensions of the averaged magnetization are
$x_m=(3-\sqrt{5})/4\approx0.191$ \cite{fisher} and $x_m^s=1/2$
\cite{mccoy2}.  The model is anisotropic at the critical point, the
dynamical exponent is $z=\infty$. More precisely the characteristic
length scale $\xi$ and the corresponding time scale $t$ are related
through:
\be
\ln t \sim \sqrt{\xi}
\label{ten}
\ee
thus the model is not conformally invariant and predictions in
eqs.\ (\ref{five}-\ref{seven}) are not expected to be valid.

In the following we briefly describe how the density profiles were
calculated.  The local magnetization $m_l$ is obtained from the
asymptotic behavior of the (imaginary) time-time correlation function
$G_l(\tau)=\langle \sigma_l^x(\tau) \sigma_l^x(0)\rangle =\sum_i\,
|\langle i|\sigma_l^x|0\rangle|^2\, \exp[-\tau(E_i-E_0)]$ where
$\langle 0|$ and $\langle i|$ denote the ground state and the $i$-th
excited state with energies $E_0$ and $E_i$, respectively. In the low
temperature (strong coupling) phase $E_1$ is asymptotically degenerate
with the ground state, thus the sum is dominated by
the first term. In the large $\tau$ limit $G_l(\tau)= m_l^2$,
therefore the local magnetization is given by \cite{remark2}
\be
m_l=\langle 1|\sigma_l^x|0\rangle~~~.
\label{twelve}
\ee

The energy-density profile is given by the ground state expectation
value $e_l=\langle 0|\sigma_l^z|0\rangle$. Since $e_l$ contains a non-singular
contribution the scaling behavior of the energy-density is more
convenient to deduce from the asymptotic form of the connected
time-time correlation function of the energy-density operator
$\sigma_l^z$. Similarly to the order-parameter the singular energy
density $\epsilon_l$ is given by a matrix-element:
\be
\epsilon_l=\langle \epsilon|\sigma_l^z|0\rangle~~~,
\label{thirt}
\ee
where $\langle \epsilon|$ denotes the first excited state, which has
non-vanishing matrix-element with the ground state.

To calculate the matrix-elements in eqs.\ (\ref{twelve}) and
(\ref{thirt}) we first, following Lieb {\it et al} \cite{lieb} and
Pfeuty \cite{pfeuty}, transform $\hat H$ into a free-fermion model.
For the fixed and free b.c.\ we study in this letter we found it most
convenient to choose the representation described in \cite{igloi},
which necessitates only the diagonalization of an
$2L\times2L$-tridiagonal matrix. From the corresponding eigenvectors
one obtains the local magnetization (\ref{twelve}) and local energy
density (\ref{thirt}) as described in \cite{berche}.

The critical properties of random Ising chains are expected to be
independent of the details of the distributions of the couplings
and/or fields. In this Letter we consider two different cases: the
binary distribution
\be
\pi(J)={1 \over 2} \delta(J-\lambda)+{1 \over 2} \delta(J-\lambda^{-1})
\;;\quad h=h_0\;,
\label{eightt}
\ee
i.e.\ $\rho(h)=\delta(h-h_0)$, and the uniform distribution:
\be
\pi(J)=\theta(1-J)  \theta(J)\;;\quad
\rho(h)=\theta(h_0-h)\theta(h)
\label{ninet}
\ee
In both cases the critical point is at $h_0=1$. All numerical data
which we present below are averaged over 50,000 samples and the
resulting statistical error is much smaller than the size of the
symbols used in the plots. Disorder-averaged quantities are denoted by
the brackets $[\ldots]_{\rm av}$.

\begin{figure}[hbt]
\epsfxsize=\columnwidth\epsfbox{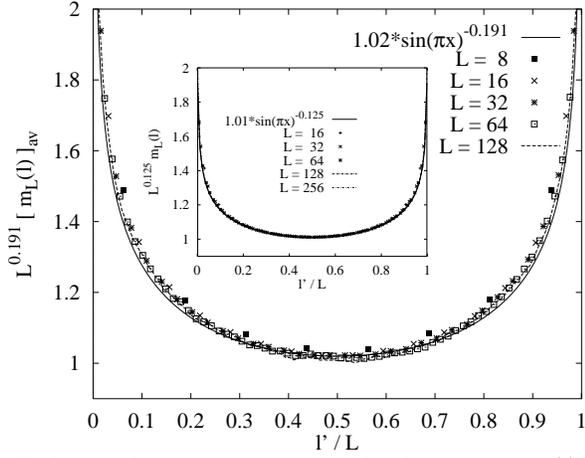}
\caption{
        Scaling plot of the magnetization profile $m_L(l)$ 
        (\protect{\ref{twelve}}) with fixed
        b.c.\ at both ends. We have shifted the site index by
        half a lattice constant and denote $l'=l-0.5$.
        The pure case is depicted in the inset,
        for which the scaling function is given by (\protect{\ref{six}})
        with $x_m^{\rm pure}=1/8$ and $G_{++}=const.$
        The main figure shows the result 
        for the binary distribution (\protect{\ref{eightt}}) with 
        $\lambda=4$. Other values of $\lambda$ as well as the uniform
        distribution yield the same quality for the data collapse,
        with different values for the non-universal prefactors but 
        identical scaling function (\protect{\ref{six}}) with
        $x_m^{\rm random}=\beta/\nu\approx0.191$ and $G_{++}=const.$
}
\end{figure}

First we study the magnetization profile of the system with fixed
b.c. at both ends of the chain. The finite size results on the pure
model, which are shown in the inset of Fig.\ 1 are in complete
agreement with the conformal prediction in (\ref{five}).  The profile
for the random chain is shown in Fig.\ 1. From the scaling plot one
can see that the Fisher-de Gennes scaling result in (\ref{one}) is
well satisfied with the conjectured value of the decay exponent
$x_m=\beta/\nu=.191$. Note that we do {\it not} use $x_m$ (as well as
later $x_m^s$) as fit paramters but fix them to the theoretically
predicted values cited above. The only fit parameter is the
non-universal prefactor $A$ in (\ref{five}). Obviously, one can very
accurately describe the finite-size data in the whole profile with the
first term of the Fourier expansion in (\ref{four}). The corrections
to the conformal result in (\ref{five}) are indeed negligible.

Next we turn to study the magnetization profiles with free-fixed
b.c. As seen on the inset of Fig.\ 2 the finite-lattice results on the
pure model perfectly coincide with the conformal prediction in
(\ref{seven}).  Results for random models are shown in Fig.\ 2.  As
one can see the numerical data collapse to a scaling function, which
can be very accurately described by a function of the form in
(\ref{seven}) with the exponents: $x_m=.191$ and
$x_m^s=\beta^s/\nu=1/2$ (again the only fit-parameter is the
non-universal prefactor $A\cdot B$ from (\ref{five}-\ref{six}).
According to Fig.\ 2 the corrections to the conformal result seem here
also to be negligible.

\begin{figure}[hbt]
\epsfxsize=\columnwidth\epsfbox{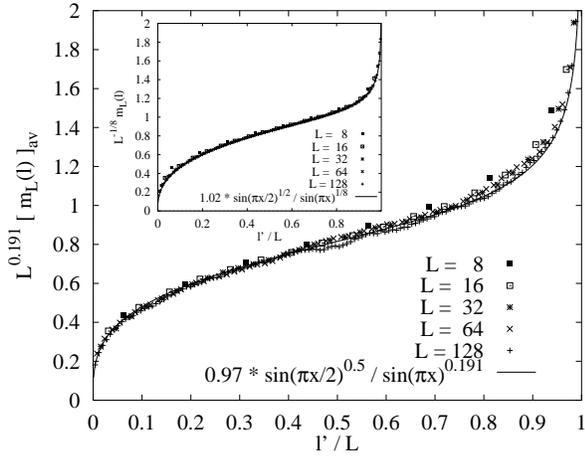}
\caption{ The same as in fig.\ 1 with fixed b.c.\ on the right end of
        the chain and free b.c.\ on the left end. The magnetization
        profile is given by (\protect{\ref{six}}) and
        (\protect{\ref{seven}}) with $x_m^s=1/2$ for the pure {\it
        and} the random case. The data shown in the main figure are
        for the uniform distribution (\protect{\ref{ninet}}).  }
\end{figure}

Due to symmetry the magnetization in a finite system with free
(non-symmetry breaking) b.c. is zero. However,
introducing an infinitesimal symmetry breaking field $h$, one can
obtain the magnetization by Yang's method \cite{yang} with the
identical result as in (\ref{twelve}). Then, as $h \to 0$ the
matrix-element in (\ref{twelve}) can be considered to define the local
magnetization in a finite system in a time scale $\tau \ll t$, where
$t$ is the relaxation time. The profile of the matrix-element in
(\ref{twelve}) can be predicted by conformal invariance
\cite{turban}. For a general local operator $\hat \Phi(l)$ the scaling
form in the strip geometry is given by \cite{turban}:
\be
\langle0|\hat \Phi(l)|\Phi\rangle \propto 
\left( {\pi \over L} \right)^{x_{\Phi}}
\left( \sin \pi{l \over L} \right)^{x_{\Phi}^s-x_{\Phi}}
\label{twentyone}
\ee
where $x_{\Phi}^s$ denotes the surface scaling dimension of $\hat
\Phi$. This expression satisfies the known scaling limits
$\langle0|\hat \Phi(1)|\Phi\rangle \sim L^{-x_{\Phi}^s}$ and
$\langle0|\hat \Phi(L/2)|\Phi\rangle\sim L^{-x_{\Phi}}$ at the surface
and in the bulk, respectively. For non-conformally invariant systems
(\ref{twentyone}) represents the first leading term of a
Fourier-expansion, as in (\ref{four}) and (\ref{five}).

\begin{figure}[hbt]
\epsfxsize=\columnwidth\epsfbox{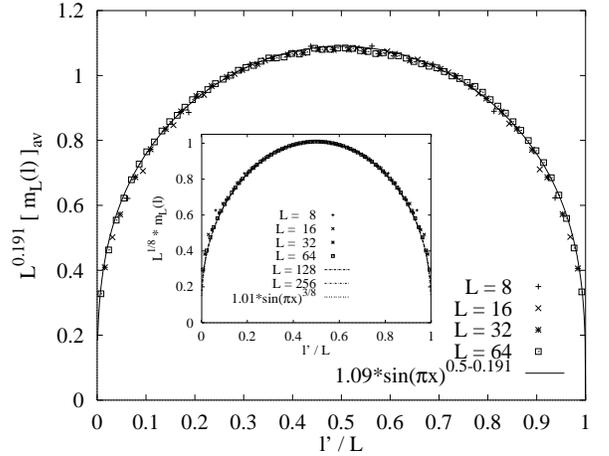}
\caption{ The same as in fig.\ 1-2 with free b.c.\ on both ends of the
        chain. The data for the random case are for a binary
        distribution with $\lambda=2$.  The magnetization profile is
        given by (\protect{\ref{twentyone}}). The data shown in the
        main figure are for the uniform distribution
        (\protect{\ref{ninet}}).  }
\end{figure}

Numerical results on the magnetization profiles with free boundary
conditions are shown on Fig.\ 3. Again the finite-size results on the
pure Ising model are in complete coincidence with the conformal
prediction in (\ref{twentyone}).  For the random case the numerical
data collapse to a scaling curve, which is very accurately described
by the conformal expression in (\ref{twentyone}) with the exponents
$x_m=.191$ and $x_m^s=1/2$. Thus again the non-conformal corrections
are very small.

Finally, we discuss the singular part of the energy density profile and
study the energy density matrix-element in (\ref{thirt}). For the pure
model one can easily evaluate $\epsilon(l)$,
which yields in the scaling limit ($l\gg 1,~L\gg 1$):
\be
\epsilon(l)={2 \over L} \sin \pi {l \over L}
\label{twentythree}
\ee
This corresponds to the conformal result in eq(\ref{twentyone}) with
$x_e=1$ and $x_e^s=2$.

In a quantum system the bulk energy-density is proportional to the
inverse relaxation time: $\epsilon \sim t^{-1}$. In the
random transverse Ising chain the scaling is anomalous as indicated in
(\ref{ten}), therefore the appropriate scaling combination is
$L^{-1/2}\ln \epsilon(l)$ instead of $L\epsilon(l)^{1/z}$ if $z$ would
be finite. In the following we study the typical energy density
$\left[\ln \epsilon(l) \right]_{\rm av}$, which after multiplication with
$L^{-1/2}$ yields a universal scaling function. The finite size data
for the random case very well satisfy the relation:
\be
\left[\ln \epsilon(l) \right]_{av} L^{1/2}=A_0+A_1 \left( {L \over \pi}
\sin\pi{l \over L} \right)^{1/2}
\label{twentyfour}
\ee
We note that this expression can also be considered as the leading
part of a Fourier expansion, where the correction terms are again very
small.

\begin{figure}[hbt]
\epsfxsize=\columnwidth\epsfbox{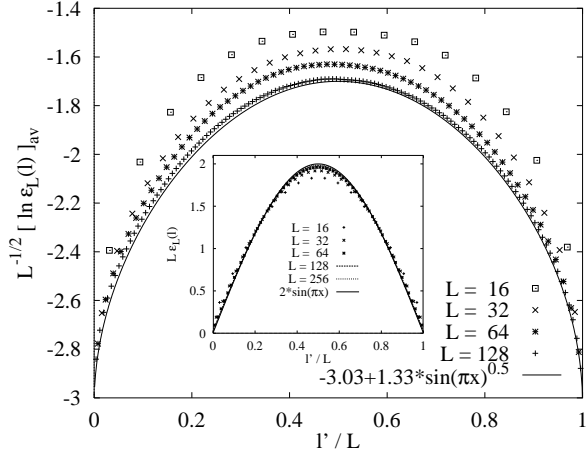}
\caption{ The energy profile $\epsilon_L(l)$
        (\protect{\ref{thirt}}) with free b.c.\ on both sides. The
        pure case is depicted in the inset, for which the scaling
        function is given by (\protect{\ref{twentythree}}). The main
        figure shows the result for the binary distribution
        (\protect{\ref{eightt}}) with $\lambda=4$. Here the scaling
        function is well described by (\protect{\ref{twentyfour}}). We
        note that the approach to the asymptotic scaling limit seems
        to be much slower than for the magnetization profiles.}
\end{figure}

To summarize we have investigated the density profiles of random
transverse-field Ising spin chains. The numerical data on rather large
systems $L \le 128$ follow scaling plots and the scaling functions can
be described very accurately by analytical expressions, which are
derived for conformally invariant systems. Since our system is not
conformally invariant there are presumably corrections. These are,
however, very small, certainly smaller than the error in our present
numerical calculation.

Generally the non-conformal corrections to the density profiles are
not small.  As an example we mention the two-dimensional aperiodically
layered Ising model \cite{igloi2}, which is somewhat related to our
problem.  When the aperiodically modulated couplings of the model
represent a marginal perturbation the system is described by a
coupling dependent dynamical exponent $z>1$ \cite{igloi}, thus the
system is not conformally invariant.  Although the aperiodic model
looks similarly to our random problem its density profiles are
completely different from the conformal results \cite{berche}. One
could speculate about the existence of some hidden symmetry which
explains the coincidence of the density profiles of the random
transverse-field Ising chain with the conformal result.

Acknowledgment: This study was partially performed during our visits
in J\"ulich and Budapest, respectively. This work has been supported
by the Hungarian National Research Fund under grants No OTKA TO12830
and OTKA TO17485.  F.\ I.\ is indebted to L. Turban for exchanging ideas
about the subject. H.\ R.'s work was supported by the Deutsche
Forschungsgemeinschaft (DFG).


\begin{references}
\bibitem{remark2}
        The fixed b.c.\ situation is formally treated by taking 
        $h_1$ and/or $h_L$ equal to zero. Then the ground state 
        is degenerate and eq.\ (9) gives the ground state expectation
        value of the magnetization.
\end{references}
\end{document}